\documentclass{svproc}
\usepackage{url}
\usepackage{xcolor}
\usepackage{graphicx}

\begin{document}
\mainmatter              
\title{A Testing Tool for IoT Systems Operating with Limited Network Connectivity}
\titlerunning{Testing Tool for IoT Systems with Limited Network Connectivity}  
%

\author{Matej Klima \and
Miroslav Bures}



\authorrunning{M.Klima and M.Bures} 
%


\institute{Dept. of Computer Science, FEE, Czech Technical University in Prague, Czechia 
\email{miroslav.bures@fel.cvut.cz}\\
\url{http://still.felk.cvut.cz}}

%

\maketitle              

\begin{abstract}
For Internet of Things (IoT) systems operating in areas with limited network connectivity, reliable and safe functionality must be ensured. This can be done using special test cases which are examining system behavior in cases of network outage and restoration. These test cases have to be optimal when approached from the testing effort viewpoint. When approached from the process viewpoint, in the sense that a business process supported by a tested system might be affected by a network outage and restoration, test cases can be automatically generated using a suitable model-based testing (MBT) technique. This technique is currently available in the open freeware Oxygen MBT tool. In this paper, we explain the principle of the technique, a process model of the tested system that may be affected by limited network connectivity, and support for this specialized MBT technique on the Oxygen platform.
\end{abstract}

\section{Introduction}

\color{blue}
Paper accepted at \textbf{WorldCist'21 - 9th World Conference on Information Systems and Technologies}, Portugal, 30-31 March to 1-2 April 2021
\newline
\newline
\textbf{http://www.worldcist.org/}
\newline
\color{black}

The reliability and security Internet of Things (IoT) systems operating in areas with limited network connectivity might be threatened by defects arising from the incorrect handling of network connectivity outages and restorations \cite{kiruthika2015software}. 

This is especially true for dynamic IoT systems, in which the connected devices move spatially. Examples of such systems are intelligent transport systems, smart cities, smart logistics systems, smart farms, dynamic sensor networks, or defense systems operating at sea or in uninhabited areas. Network coverage can be limited for various reasons. The system may operate in uninhabited or rural areas where network coverage is naturally weaker, or in urban areas where network connections can be interrupted (by tunnels, for example). 

Regardless of these possible network connectivity limits, an IoT system under test (SUT) must run consistently and be free of faults or interruptions owing to weak or broken network connections. In such a situation, the users might accept the deterministic temporary outage in SUT functionality when they are given proper feedback. However, the users are less likely to accept nondeterministic random behaviour of the system caused by network outages, which would lead to inconsistencies in data processing or critical defects, which can cause harm to the businesses supported by the SUT or even to the system operators or users.

Hence, when network connectivity in an IoT system is interrupted and restored, this system must be systematically and effectively tested to ensure its functionality. In this paper, we describe a specialized MBT technique to design such test cases for processes in an IoT SUT that might be affected by limited network connectivity. This technique is currently implemented in the open freeware the Oxygen MBT tool, which is also introduced in this paper.


\section{Principle of the Technique}

In testing the SUT functionality in a situation of possible limited network connectivity, we are principally interested in two situations.

The first is \textbf{an interruption of a network connection}. In such a situation, we are interested in testing the following questions: (1) Will the system correctly inform the user (where relevant)? (2) If data are collected, are they stored in a cache, or are they lost? If they are stored in a cache, how long the cache can hold the collected data? (3) If a device or module of the system that is affected by a limited network connection accepts signals or commands from other devices or parts of the system, are these parts notified that the device or module is offline? (4) If the data or signals are processed in transactions, does the SUT maintain this transactional behavior when the network connection is interrupted?


The second situation is \textbf{the restoration of the network connection}. Here, the following cases might be tested to verify the SUT behavior: (1) If relevant, is the SUT user or operator notified that the connectivity had been restored? (2) If a device or part of the SUT uses caching to overcome network connectivity outages, is the cached content correctly transmitted to the respective SUT parts? (3) Has the consistency of the stored data been maintained? (4) If there are cached transactions, are they finished correctly? Have the logical order or required timing of their steps been maintained? (5) Is the performance of the SUT not (unacceptably) affected by the return of the device or module to online mode?


These examples may not be a complete list of situations that require testing in relation to limited network connections; additional situations may be relevant regarding the specific features of particular IoT systems.

The principle of the test cases we are generating in the proposed technique is accurately simulating the situation of a network outage and its restoration. These test cases are generated optimally having maximal probability to detect the relevant defects for minimal testing costs.

\subsection{Model of the Problem}

\begin{figure}
\includegraphics[width=\textwidth]{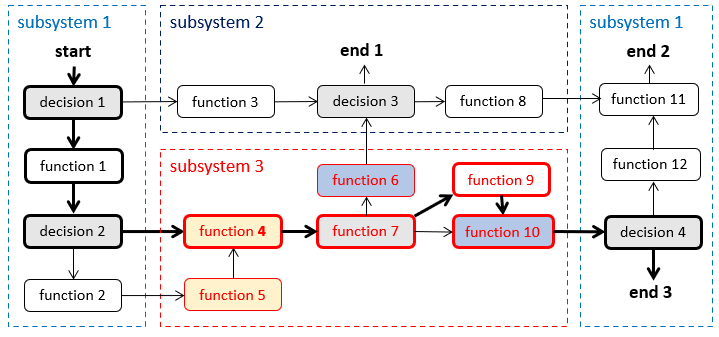}
\caption{Model of the problem SUT process affected by limited network connectivity}  \label{fig:model}
\end{figure}

In the proposed technique and its support in the Oxygen tool, we model the SUT functionality from a process viewpoint. The SUT model based on the Unified Modeling Language (UML) activity diagram simplified to a directed graph, which we illustrate in Figure \ref{fig:model}. 


The nodes of the graph represent actions, functions, or decision points of a process implemented in an SUT. The edges represent transitions in the modeling process. The decision point is a node with an outgoing degree larger than 1, whereas an action or function is a node with outgoing degree of 1. As shown in Figure \ref{fig:model}, the decision points are distinguished by a grey background.

The actions, functions, or decision points can be implemented in various connected devices of an IoT SUT, and some can be affected by limited network connectivity. To illustrate this fact, we set a \textit{network outage probability} for each element of the model. As shown in Figure \ref{fig:model}, three subsystems cooperate. Subsystem 1 handles a process flow to subsystems 2 and 3. After the completion of the sub-processes in these subsystems, the process continues in subsystem 1. In the example, subsystem 3 can be affected by limited network connectivity, whereas subsystems 1 and 2 are connected to a stable network. 

To generate the test cases, we set a \textit{network outage probability threshold}. The elements of the model having the network outage probability greater than this threshold form the \textit{limited connectivity zones (LCZs)} in the model. In the example given in Figure \ref{fig:model}, only one LCZ is present, marked by red borderline of nodes; however, more LCZs can exist in the model. 

Each LCZ has a set of entry and exit nodes. An \textbf{entry node} represents an action, function or decision point in an SUT process in which the network connectivity is disrupted. An \textit{exit node} represents an action, function, or decision point in an SUT process in which the network connectivity is restored. Through an entry node, the process flow can enter an LCZ; the process flow leaves an LCZ through an exit node. As shown in Figure \ref{fig:model}, the entry nodes are marked by a yellow background and the exit nodes by a blue background.

\subsection{Test Case and Test Coverage Criteria}
\label{subsec:coverage_criteria}

Using the SUT process model described in the previous section, we can generate the test cases. Specifically, a test case is a path in the SUT model from the process start to any of its ends that tours an entry node and is followed later in the path by an exit node (i.e., a test case enters and exits an LCZ). One test case can visit more LCZs if they are present in the SUT model. In  Figure \ref{fig:model}, a test case is depicted by the bold arrows and rectangles.

The test case generation strategy implemented in the Oxygen module (see Section \ref{sec:oxygen}) generates a set of such test cases with the goal of minimizing the total number of steps.


In this process, two test coverage levels are supported. In \textit{Each Border Node Once}, each of the entry and exit nodes must be present at least once in the test cases. In  \textit{All Combinations of Border Nodes}, all possible pairs of entry nodes with exit nodes must be toured by the test cases. The latter test coverage level leads to a test set with a higher number of test steps, which also entails a potentially higher probability of detecting a defect.

\section{Tool Support: Oxygen Module}
\label{sec:oxygen}

The presented MBT technique is implemented as a specialized module of the open freeware MBT platform Oxygen\footnote{http://still.felk.cvut.cz/download/oxygen-iot.zip}, developed and issued by our research group. The Oxygen platform was developed in Java and is available to use as an executable JAR file, requiring Java 1.8. 

\begin{figure}
\includegraphics[width=\textwidth]{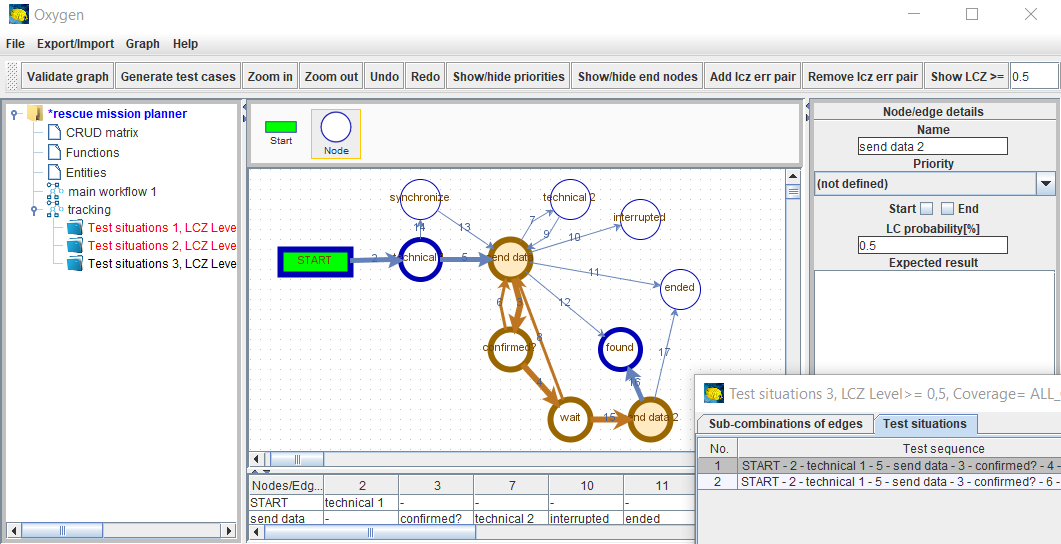}
\caption{Editor of SUT process in the Oxygen and visualization of LCZs.} \label{fig:oxygen_editor}
\end{figure}

The model of the SUT process that may be affected by limited network connectivity can be created in the Oxygen user interface (an example is given in Figure \ref{fig:oxygen_editor}). For each part of the model, the network outage probability can be defined and the LCZs can be visualized for a given network outage probability threshold (subgraph depicted in brown in Figure \ref{fig:oxygen_editor}). The entry and exit nodes of the LCZ are depicted by a light brown background.

To generate the test cases, the user selects the test coverage criteria (see Section \ref{subsec:coverage_criteria}) from the dialogue, as well as network outage probability threshold, thereby forming the LCZs.

To generate the test cases, the portfolio strategy composing of three algorithms is available in the project. These algorithms are:

\begin{itemize}
    \item \textbf{Shortest Path Composition Algorithm} that constructs the test cases, firstly, by searching for the shortest paths from the entry nodes to the exit nodes in the LCZs. Secondly, it constructs the proper test cases by creating the paths from the start node to one of the end nodes, using, if possible, the paths found in the previous step.
    \item \textbf{Ant Colony Optimization (ACO) Based Algorithm} is the implementation of the original ACO algorithm, introduced by Dorigo \cite{dorigo1992thesis}. Originating in the start node, the ants are exploring the model, preferring the paths made of nodes that lead to the undiscovered LCZs. During this traversal, each ant leaves its pheromone behind, which attracts the other ants. The best path found by the ants is considered then as a test case. This process is executed repeatedly until the generated set of test cases doesn't satisfy the coverage criterion.
    \item \textbf{Enforced Prime Paths Algorithm}, which builds on the previous path-based testing concepts. This algorithm creates a set of test requirements to satisfy the given test coverage criteria and then employs the matching-based prefix graph algorithm for prime paths search \cite{li2012better} to compose the test cases.
\end{itemize}
\color{black}
The generated test cases appear in the left project tree and are saved with the project. When the user updates the SUT model, the previous test cases may become obsolete, which is illustrated in red in the project tree. 

The test cases can also be visualized in the SUT model (Figure \ref{fig:oxygen_editor} shows an example of a pop-up window with the generated test cases in the bottom right corner of the screen and the test cases visualized in the model editor as bold nodes and edges). In addition, the test cases can be exported in open formats based on CSV, XML, and JSON and can be further imported to a test automation framework or to a test management tool, where they can be further elaborated to detailed tests.

The Oxygen tool with the support of the technique described in this paper is freeware.

\section{Related Work and Discussion}

The technique proposed in this paper is compared with the current path-based testing techniques. In this field, an SUT model is based on a directed graph \cite{bures2019employment} and several algorithms are available to generate the test cases \cite{bures2019employment,li2012better,arora2017synthesizing,anand2013orchestrated}. However, no technique directly supports the requirements of our concept of limited network connectivity testing; specifically, that a particular node must be toured, and this node has to be followed by another exactly defined node later in the path \cite{anand2013orchestrated}.
This requirement, together with the goal of minimizing the test set, makes our proposed technique novel in the field.

Indirectly, the problem can be solved using the test requirements concept. A test requirement is a path in the SUT model that must be present in the test cases \cite{li2012better,anand2013orchestrated}. To solve the problem using this approach, we can define a path across the LCZ from its entry node to its exit node as a test requirement. Thus, the algorithm generating the test cases as required in our technique can be built upon an algorithm generating the path-based test cases that accepts an SUT model and a set of test requirements. However, extensions for such an algorithm must be created.

From alternative techniques to test the behaviour of an IoT SUT under limited connectivity, the current work is focused mainly on the network level; for instance \cite{rudevs2018towards,white2017quality}. In addition, such work spans from network reliability testing to quality of service assessments \cite{white2017quality,matz2020systematic}. From this viewpoint, most work focuses on the reliability of the network itself, rather the systematic testing of higher levels of an SUT operating under limited network connectivity.

From a conceptual viewpoint, the dynamic nature of the problem of possible limited network connectivity during SUT operation might point future research directions towards adding more agility to the approach as well. In this sense, dynamic construction of the SUT model and its subsequent combination with model-based test case generation, as successfully tried with web-based software systems \cite{bures2018tapir,frajtak2017exploratory,bures2016smartdriver}, might be a perspective method as well in the IoT systems. Such an approach has to be examined further.

\section{Conclusion}

In the paper, we present a module of the open freeware MBT platform Oxygen, which facilitates the generation of effective test cases for IoT systems operating with limited network connectivity. In the proposed technique, the SUT processes and models the points that may be affected by network outages, and the test cases are automatically generated. The proposed model is based on a directed graph and employs additional metadata regarding the probability of a network connectivity outage.


Two test coverage criteria levels are supported in the current version of the technique, suitable for low-intensity tests and thorough high-intensity tests. High-intensity tests are resulting in a higher number of more prolonged test cases and are suitable for critical parts of SUT or functionality that repeat frequently.

The proposed method, which focuses on a process-oriented view, can serve as a complement to the current network layer-focused techniques commonly operating on lower layers of an IoT system.

\section*{Acknowledgements}


This research is conducted as a part of the project TACR TH02010296 Quality Assurance System for the Internet of Things Technology. The authors acknowledge the support of the OP VVV funded project CZ.02.1.01/0.0/0.0/16\_019 /0000765 “Research Center for Informatics”.

\bibliographystyle{splncs03}
\bibliography{references}

\end{document}